\documentclass[12pt,preprint]{aastex}
\usepackage{graphics}

\begin{document}
\slugcomment{Submitted to ApJ}

\title{No periodicity revealed for an ``eclipsing'' ultraluminous supersoft
X-ray source in M81}

\author{Ji-Feng Liu}

\affil{Harvard-Smithsonian Center for Astrophysics}

\begin{abstract}

Luminous supersoft X-ray sources found in the Milky Way and Magellanic Clouds
are likely white dwarfs that steadily or cyclically burn accreted matter on
their surface, which are promising type Ia supernova progenitors.  Observations
of distant galaxies with Chandra and XMM-Newton have revealed supersoft sources
that are generally hotter and more luminous, including some ultraluminous
supersoft sources (ULSs) that are possibly intermediate mass black holes of a
few thousand solar masses.  In this paper we report our X-ray spectral and
timing analysis for M81-ULS1, an ultraluminous supersoft source in the nearby
spiral galaxy M81.  M81-ULS1 has been persistently supersoft  in 17 Chandra
ACIS observations spanning six years, and its spectrum can be described by
either a $kT_{bb}\approx70$ eV blackbody for a $\sim1.2M_\odot$ white dwarf, or
a $kT_{in} \approx 80$ eV multicolor accretion disk for a $\gtrsim10^3M_\odot$
intermediate mass black hole.  In two observations, the light curves exhibited
dramatic flux drop/rise on time scales of $10^3$ seconds, reminiscent of
eclipse ingress/egress in eclipsing X-ray binaries.  However, the exhaustive
search for periodicity in the reasonable range of 50 ksec to 50 days  failed to
reveal an orbital period.  The failure to reveal any periodicity is consistent
with the long period ($\ge30$ yrs) predicted for this system given the optical
identification of the secondary with an asymptotic giant star. Also, the
eclipse-like dramatic flux changes in hours are hard to explain under the white
dwarf model, but can in principle be explained by disk temperature changes
induced by accretion rate variations under the intermediate mass black hole
model.

\end{abstract}


\section{INTRODUCTION}

Luminous supersoft X-ray sources were discovered with the Einstein Observatory
and were established as an important new class of X-ray binaries based on ROSAT
observations of 18 sources in the Milky Way and the Magellanic Clouds (Greiner
1996; Kahabka \& van den Heuvel 1997 and reference therein).
These sources have extremely soft spectra with equivalent blackbody
temperatures of 15-80 eV and are highly luminous with bolometric luminosities
of $10^{35} - 10^{38}$ erg s$^{-1}$.
They are thought to be white dwarfs that are steadily or cyclically burning
hydrogen accreted onto the surface. Remarkably, the accretion rates must be in
a narrow range of $1-4\times10^{-7} M_\odot yr^{-1}$, and the resultant
luminosities are below $10^{39}$ erg s$^{-1}$ as observed.
White dwarfs with steady nuclear burning are promising Type Ia supernova
progenitors because, unlike explosive nova events, they can retain the accreted
matter and their mass can increase till it approaches the Chandrasekhar limit.

More and more supersoft sources have been discovered in distant galaxies with
the advent of the Chandra and XMM-Newton X-ray observatories.
These distant supersoft sources, as compared to the well-studied ones in the
Milky Way and Magellanic Clouds, are generally hotter, more luminous and
largely associated with spiral arms, suggesting they are young and massive
systems (Di Stefano \& Kong 2004).
Some have bolometric luminosities, as derived from blackbody models, far above
the Eddington luminosity ($L_{Edd} = 4\pi G M m_p c/\sigma_T \approx
1.3\times10^{38} (M/M_\odot)$ erg s$^{-1}$) for white dwarfs, which we call
ultraluminous supersoft sources (ULS).
One prototype ULS is M101-ULX1, which showed supersoft spectra of $50\sim100$
eV,  0.3-7 keV luminosities of $3\times10^{40}$ erg s$^{-1}$ and  bolometric
luminosities of $\sim 10^{41}$ erg s$^{-1}$ during its 2004 outburst (Kong et
al. 2004).

While the supersoft spectra of ULSs can be explained by a white dwarf burning
accreted materials on its surface, the white dwarf model can not explain the
extremely super-Eddington luminosities.
On the other hand, an intermediate mass black hole (IMBH) of $\gtrsim 10^3
M_\odot$, as Kong et al. suggested, can naturally explain both the high
bolometric luminosities and the supersoft spectrum given the scaling relation
$T_{disk} \propto M^{-1/4}$ between the black hole mass and the accretion disk
temperature.
In the case of M101-ULX1, The IMBH nature is further evidenced by the observed
spectral state transitions, which are reminiscent of the high-soft to low-hard
transitions for stellar black hole binaries.
Such IMBHs  may have played an important role in the hierarchical merging
scenario of galaxy formation (Madau \& Rees 2001), and are under hot pursuit by
workers in the fields of massive star clusters and of ultraluminous X-ray
sources (Miller \& Colbert 2004).

Swartz et al. (2002) observed the nearby spiral galaxy M81 with Chandra ACIS,
and discovered a bright supersoft source in the bulge.
% at R.A.=09:55:42.2, Decl.=69:03:36.5.
%
Its supersoft spectrum can be fitted by a blackbody model with a temperature of
$\sim80$ eV and a bolometric luminosity of $\sim10^{39}$ erg s$^{-1}$. This
source is thus a ULS, which we designate as M81-ULS1 in this paper.
Given the high luminosity, Swartz et al. suggested that it could be a white
dwarf burning Helium instead of Hydrogen, or an intermediate mass black hole
(IMBH) of $\sim1200/(cosi)^{1/2} M_\odot$ with $i$ as the inclination angle.
A sudden flux drop by a factor of $\ge10$, Swartz et al. reported, occurred in
less than one hour in the observation ObsID 390.
This resembles the ingress of an eclipsing X-ray binary with a short period of
several to several tens of hours, comparable to the periods for some canonical
supersoft sources (Greiner 1996).

Here we report the follow-up X-ray study of M81-ULS1 with the 17 Chandra ACIS
observations accumulated since its discovery.
In \S2 we describe the reduction of the observations and the spectral and
temporal analyses of ULS1.
Presented in \S3 are the searches for possible periods for ULS1, which fail to
obtain any period. 
The results are summarized and discussed in \S4. 
The distance of M81 is taken to be 3.63 Mpc ($\mu=27.8$ mag; Freedman et al. 1994) in this work.

\section{ANALYSIS OF X-RAY DATA}

There have been 17 Chandra observations of M81 nuclear region, where ULS1
resides. 
As listed in Table 1, these include two observations (ObsID 390, 2.4 ksec,
2000-03-21; ObsID 735, 50 ksec, 2002-05-07) analyzed by Swartz et al. (2002),
and 15 monitoring observations of 10$\sim$12 ksec each from 2005-05-26 to
2005-07-06 with an average separation of 3 days (PI: D. Pooley). 
All these observations were analyzed uniformly with CIAO 3.4. Point sources
were detected with WAVDETECT on the individual and combined Chandra images. 
ULS1, at R.A.=09:55:42.2, Decl.=69:03:36.5, appeared on the S3 chip in all
observations, and is about $0.^\prime8$ away from the M81 nucleus. 
Its photon lists for timing and spectral analysis were extracted from the
source eclipses enclosing 95\% of the total photons as reported by WAVDETECT.

The light curve of ULS1 from the 17 observations over six years is shown in
Figure 1. 
The count rates in these observations were corrected for deadtime and
vignetting, which are usually less than 10\% as listed in Table 1. 
The 2005 monitoring observations caught the source in a relatively low
intensity state ($\le$ 0.01 count/sec) on 7 occasions, a high state ($>0.01$
count/sec) on 7 occasions, and a low to high transition in ObsID 5944. These
high states, intermittent with 2 low states, clustered in about a month and
appeared as a period of super activity out of low states.  
It is, however, not clear whether the low states or the high states represent
the ``normal'' condition of the source due to the limited number of
observations.
Inspection of the individual light curves shows that the flux was quite steady
during most observations.
However, sudden flux drop and rise on time scales of $10^3$ sec, as shown in
the Figure 1 inserts, were present in ObsID 390 and 5944, resembling
ingress/egress in eclipsing X-ray binaries. 
In addition, the light curve in Obs ID 5940 exhibited a few bins above 0.01
ct/sec before its gradual decrease to below 0.01 ct/sec, and could have caught
the tail of an ingress.
The binary properties will be significantly constrained if an orbital period
can be found, and the next section is devoted completely to searches for such a
period.

ULS1 has been persistently supersoft in all 17 observations spanning six years.
As shown in Table 1, the majority of the photons are in the soft band of
0.3-1.1 keV. 
For the 14 observations with $\ge$30 photons, the numbers of hard photons above
1.1 keV are only $\le$2\% of the total photons in 12 observations; the
percentages are about 5\% and 15\% for ObsID 5944 and ObsID 390, which happen
to be the two observations that show ``eclipses'' in their light curves. 
While photons are expected to harden in the eclipses, 12 out of the 13 photons
in the ``eclipse'' of ObsID 390 (i.e., in the low state) are below 1.1 keV,
thus the photon energy distribution is no harder than for photons out of
``eclipse'' (ie., in the high state).
Similarly for ObsID 5944, 14  out of the 15 photons in the ``eclipse'' are
below 1.1 keV.
Further comparison between the energy distributions for photons in the
low-state observations and for photons in the high-state observations reveals
no significant difference.
Spectra with $\ge$200 photons are fitted to absorbed blackbody models within
0.3-2 keV, and the results are listed in Table 1. The blackbody fits are quite
acceptable, with temperatures clustered around 75 eV with a 11 eV standard
deviation. The X-ray luminosities in 0.3-2 keV are in the range of 1.8-2.8
$\times10^{38}$ erg s$^{-1}$, and the bolometric luminosities are in the range
of 1.9-7.2 $\times10^{39}$ erg s$^{-1}$.
The neutral hydrogen column densities appear to cluster around two values,
$9\times10^{20}$ cm$^{-2}$ and $18\times10^{20}$ cm$^{-2}$, but there are no
apparent correlations with temperatures or count rates.

The spectral parameters are best constrained with the longest observation ObsID
735.
This spectrum, as plotted in Figure 2, 
%contains $\ge$3000 photons, and 
can be well fitted in the band of 0.3-2 keV by an absorbed black body model,
% for a nuclear burning white dwarf,
with $\chi^2_\nu/dof = 1.196/38$, $n_H = 8.6\pm0.9\times10^{20}$ cm$^{-2}$, $kT
= 73\pm1.5$ eV, $L_X(0.3-2keV) = 3.2\times10^{38}$ erg s$^{-1}$, and $L_{bol} =
2.5\times10^{39}$ erg s$^{-1}$.
The spectral fit shows residuals in the two bins above 1 keV, suggesting
presence of a hard component barely detected by the ACIS observation.
Note that our spectral fits are slightly different from Swartz et al.  (2002)
because of different choices of energy bands and changes in the calibration
files over the years.
Like in canonical supersoft sources, the blackbody emission can come from the
nuclear burning on the surface of a white dwarf.
For such a system, the surface temperature 
%and the bolometric luminosity
increases monotonically with the white dwarf mass as illustrated in Iben (1982;
Figure 2). 
The best-fit blackbody temperature would imply a white dwarf mass of
$\gtrsim1.2M_\odot$.

The spectrum can be slightly better fitted by a multicolor disk  model, with
$\chi^2_\nu/dof = 1.073/38$, $n_H = 10.5\pm0.9\times10^{20}$ cm$^{-2}$,
$kT_{in} = 83\pm1.4$ eV and $L_X(0.3-2keV) = 3.2\times10^{38}$ erg s$^{-1}$.
As shown in Figure 2, the difference between the blackbody and multicolor disk
models is hardly discernible below 0.8 keV. Above 0.8 keV, the diskbb model
begins to predict slightly larger count rates, by 10\% at 1 keV.
As in the blackbody fit, an insignificant hard component is revealed only from
the fitting residuals in two bins above 1 keV.
For a thermal accretion disk, the black hole mass can be expressed as $M
\approx 10 ({\eta \over 0.1}) ({\xi \kappa^2 \over 1.19}) ({L_{disk} \over
5\times 10^{38}})^{1/2} ({kT_{in} \over 1 keV})^{-2} M_\odot$, with $\eta$ as
the radiative efficiency, $\kappa$ as the hardening factor, and $\xi$ as
introduced to normalize the bolometric luminosity (Soria 2007 and references
therein).
The best fit inner edge temperature and the model luminosity would imply a
black hole mass $M \approx 3000 M_\odot$ with all factors set to their default
values.
Alternatively, the model normalization ($\equiv (R_{in}/D_{10})^2\cos i$;
$D_{10}$ is the distance in 10 kpc) gives the inner disk radius $R_{in} \approx
24500/(cosi)^{1/2}$ km; assuming $R_{in}$ corresponds to the last stable orbit,
this implies an IMBH of $\sim2700/(cosi)^{1/2} M_\odot$, quite consistent with
the estimate from the disk temperature.

\section{SEARCHES FOR PERIODICITY}

The light curves of ULS1 during ObsIDs 390/5944 exhibited ``eclipses''
suggestive of an eclipsing binary.
To search for possible periodicity from the 17 observations, we binned the
individual light curves to have an average of $\ge15$ counts per bin for
low-state ($\le0.01$ ct/sec) observations, to have a total of 30 bins for
high-state ($\ge0.01$ ct/sec) observations, and adjusted the bins to exhibit
the sudden flux drop/rise in ObsIDs 390/5944.
If the variations are mainly caused by the the donor eclipsing the accretor,
and the bins in the low states of ObsIDs 390/5944 and in 7 low-state
observations are in the eclipse, the low-state bins should cluster together to
form only one eclipse when individual light curves are folded with the true
period.
Were ULS1 an eclipsing binary, the period should be longer than the longest
observation ObsID 735 (50 ksec), during which the flux was steadily at a high
state of 0.07 ct/sec. Also, the period should be shorter than $\le$50 days
given the alternating low/high states during the 2005 monitoring observations.

Two techniques were applied on the 17 binned light curves to search for
possible periods.
The first was the Lomb-Scargle method (Lomb 1976; Scargle 1982), which was
devised to work on unevenly spaced data as in this case.
This method calculates the Lomb periodogram for a range of trial periods and
quantifies the significance as the probability for the periods to originate
from a purely random signal.
The Lomb periodogram for ULS1 is plotted in Figure 3, which shows a multitude
of significant candidate periods with the probability from randomness smaller
than $10^{-10}$.
These candidate periods cluster around $10-10^2$ hours and $10^3$ hours, which
is about the length of the 2005 monitoring observations.
To check how the time windows of the observations affect the results, we
computed the Lomb periodogram for Monte Carlo permutations of the observed
light curves (i.e., re-assigned the count rate for one time bin to another bin
randomly). 
As shown in Figure 3, the power for the Monte Carlo permutations is far less
than for the observations, indicating that the resulted candidate periods come
from the source variability rather than the time windows.

We also applied  the phase dispersion minimization method as demonstrated in
Stellingwerf (1978).
For each trial period, this method folds the light curves and computes the
dispersion of observations in phase bins. 
The phase dispersion was calculated for a period range of 15-1200 hours, and
the result is plotted in Figure 4.
A multitude of candidate periods are revealed as deep minima of $\Theta$, a
measure of the phase dispersion.
Given the number of data points in this calculation, $\Theta = 0.66$ corresponds
to the probability of $10^{-6}$ for the candidate period and associated phase
dispersion to come from random fluctuations.
Similar to the experiments for the Lomb-Scargle method, we find that the phase
dispersion for the Monte Carlo permutated data is much higher than for the
observed data, indicating that the resulted candidate periods come from the
source variability rather than the time windows.

The reported significance of the candidate periods, however, is only testing
the probability for the power or phase dispersion to come from random
fluctuations. 
The small probability only negates their coming from random fluctuations, but
is not always a guarantee that they come from truly periodic variations.
To check the authenticity of these candidate periods, we utilize the criterion
that, as naturally expected, there should be only one eclipse in the folded
light curve.
Eclipses are defined as dips of one or more low-state bins enclosed by
high-state bins in the folded light curve. Technically, we assign 0/1 to
low-state/high-state bins, and count an eclipse when a jump from 0 to 1 is
detected in the folded light curve.
This eclipse counting method is conceptually similar to the string-length
method (Dworetsky 1983), but is computationally more efficient.
We have computed the numbers of eclipses for several tens of significant
candidate periods and visually checked their folded light curves.
Interestingly, none of them could give only one eclipse in their folded light
curves.
Plotted in Figure 3 as an example is the folded light curve with four
``eclipses'' for the candidate period of 96.291 hours, the most significant
from the phase dispersion minimization method.
Apparently, such candidate periods are not qualified as the true orbital period
for ULS1.

The failure to qualify these significant candidate periods led to the suspicion
that there was no binary periodicity in these observations at all.
However, a true period might have escaped the detection of the two employed
methods for some reasons, we thus tried to examine a continuous range of
periods from 50 ksec to 50 days with the one-eclipse criterion in search for a
true period.
First, periods should be excluded if they make a high-state light curve and a
low-state light curve overlapping in phase.
Assume a low state curve from $t^l_0$ to $t^l_1$ in time trails a high state
curve from $t^h_0$ to $t^h_1$. The shortest period is $P_{min} =
(t^h_1-t^h_0)+(t^l_1-t^l_0)$ to avoid overlapping in phase, and the maximum
number of periods between the two curves would be 
$N_{Max} = (t^l_1-t^h_0)/P_{min}$. 
%$N_{Max} = int[(t^l_1-t^h_0)/P_{min}]$. 
%
To avoid overlapping after subtracting $n (\le N_{Max}) $ periods ($P$) from
the trailing light curve, periods should not satisfy \\
\indent \ \   $(t^h_0+nP) \le t^l_0 \le (t^h_1+nP)$, \\
\indent or  $(t^h_0+nP) \le t^l_1 \le (t^h_1+nP)$, \\
\indent or  $t^l_0 \le (t^h_0+nP) \le t^l_1$, \\
\indent or  $t^l_0 \le (t^h_1+nP) \le t^l_1$. \\
These requirements lead to forbidden periods in ranges \\
\indent \ \ \    ${t^l_0 - t^h_1 \over n} \le P \le {t^l_0 - t^h_0 \over n}$,
\\
\indent and ${t^l_1 - t^h_1 \over n} \le P \le {t^l_1 - t^h_0 \over n}$, \\
\indent and ${t^l_0 - t^h_0 \over n} \le P \le {t^l_1 - t^h_0 \over n}$, \\
\indent and ${t^l_0 - t^h_1 \over n} \le P \le {t^l_1 - t^h_1 \over n}$, \\
which can be reduced to 
${t^l_0 - t^h_1 \over n} \le P \le {t^l_1 - t^h_0 \over n}$.
In practice, we exclude periods that satisfy 
${t^l_0 - t^h_1 + \delta \over n} \le P \le {t^l_1 - t^h_0 - \delta \over n}$ 
for $1 \le n \le N_{Max}$, with $\delta$ allowing for slight overlapping of
low/high states.

The forbidden period ranges were computed for each combination of a high-state
light curve and a low-state light curve.
The light curves for ObsIDs 390/5944 were split into two separate light curves,
a high-state one and a low-state one.
The allowed period ranges  was obtained by excluding all the forbidden period
ranges from the range of 50 ksec to 50 days.
%
%Individual light curves were folded with all periods in the allowed period
%ranges.
%
The continuous intervals of allowed periods was discretized with the period
resolution set to $\Delta P = {P^2 \over 360\Delta T}$. 
Under this period increment, the relative phase changes $\Delta \phi = {\Delta
T \over P} {\Delta P \over P}$  for the bins spanning $\Delta T$ (about six
years) do not exceed $1^\circ/2\pi$, a tiny amount as compared to the width of
possible eclipses.
Such discretization makes the search for periodicity exhaustive, and guarantees
not to miss the true period if it exists.
This exhaustive search, however, failed to find any period that gave only one
eclipse in the folded light curve; in fact, all discretized periods gave three
or more eclipses as shown in Figure 6.

An assumption for the periodicity search was that all low-state bins were
caused by eclipsing. 
If some low-state bins were caused by reasons other than eclipsing, they might
have appeared as separate eclipses in addition to the true eclipse, thus
disqualifying some trial periods as the true period.
Since all discretized periods gave three or more ``eclipses'', at least two out
of the nine low-state observations (including ObsID 390/5944) should be
considered as out-of-eclipse to possibly obtain a folded light curve with only
one ``eclipse''.
The continuous period range of 50 ksec to 50 days was tested with different
combinations of two low-state observations excluded, and two combinations
%ObsIDs 5935/5944 and ObsIDs 5940/5944, 
were found to result in viable periods
with only one ``eclipse'' in the folded light curves (Figure 7).
The first combination of ObsIDs 5935/5944 led to viable periods around 96.7
hours, but the folded light curves showed an unreasonably wide eclipse
($\ge140^\circ$). 
The second combination of ObsIDs 5940/5944 led to eight intervals of viable
periods evenly spaced  in the range of 30-35 days, reflective of the duration
of the super activities (with the two intervening low-state observations ObsIDs
5940/5944 excluded) in the 2005 monitoring observations.
%

%about processing disk: no power to constrain this

\section{DISCUSSION}

Six years of Chandra ACIS observations have revealed the ultraluminous
supersoft source M81-ULS1 as persistently supersoft despite the dramatic flux
changes.
The highest quality spectrum for M81-ULS1 can be well fitted by a multicolor
disk around an IMBH of $\gtrsim10^3 M_\odot$.
The spectrum can also be well fitted by a blackbody model as for the nuclear
burning on the surface of a white dwarf of $\gtrsim1.2M_\odot$, but the
bolometric luminosity is 10 times higher than the Eddington luminosity for
white dwarfs steadily burning hydrogen on the surface, and 5 times higher than
for white dwarfs steadily burning helium (Iben \& Tutukov 1989).
The super-Eddington problem, however, may be solved by adopting more
sophisticated white dwarf atmosphere models, experiments with which showed that
they could reduce the bolometric luminosities by a factor of ten (van Teeseling
et al.  1996).
To summarize, the X-ray spectrum is consistent with M81-ULS1 being either an
IMBH of $\gtrsim10^3 M_\odot$ or a massive white dwarf.
X-ray observations thus lack the definitive power to distinguish between an
IMBH accretor or a white dwarf accretor, and observations in lower energies are
needed to understand the nature of the accretor.

The accretion disk should behave quite differently  in presence of different
accretors for the same observed X-ray luminosities.
If the accretor is a white dwarf, the accretion disk, illuminated by the soft
X-ray emission from the white dwarf surface, will flare up geometrically,
intercept a large fraction of the X-ray emission and re-emit in lower energies
(Popham \& Di Stefano 1996).
The emergent spectrum will take a form of $\nu^{-1}$ in the optical (FKR 2002).
If the accretor is an IMBH and thus the high X-ray luminosity is less than a
few percents of the Eddington luminosity, the accretion disc will be a standard
geometrically-thin optically-thick one (Shafee et al. 2007).  
The spectrum of such a thin thermal disk, conventionally described by a
multi-color disc model (Mitsuda et al. 1984), is a powerlaw of $F_\nu \propto
\nu^{1/3}$ in the optical.
Such a difference may be picked up by optical and infrared observations. 
Indeed, optical studies of M81-ULS1 have identified M81-ULS1 as a bright
point-like object, the spectral energy distribution of which can be decomposed
into an AGB star as the companion and a blue component (Liu \& Di Stefano
2007).
The blue component is consistent with a $F_\nu \propto \nu^{1/3}$ powerlaw as
from the accretion disk expected for an IMBH, but not consistent with a $F_\nu
\propto \nu^{-1}$ powerlaw expected for a white dwarf.
This is strong observational evidence that M81-ULS1 is an IMBH instead of a
white dwarf.

% - this failure to detect periods is consistent with its optical study

The light curves of ULS1 exhibited in two observations sudden flux drop/rise
%suggestive of periodicity due to eclipsing.
that resembled the eclipse ingress/egress for eclipsing X-ray binaries,
suggestive of periodicity due to eclipsing.
This tempted us to search for periodicity from the 17 observations for ULS1.
The candidate periods obtained with the Lomb-Scargle and phase dispersion
minimization methods, however, are
all false periods because they do not satisfy the criterion that a true period
should have only one ``eclipse'' in their folded light curves.
Further examinations on possible periods in a continuous range of 50 ksec to 45
days, discretized in an exhaustive way, also fail to find any true period.
We conclude that no periodicity is revealed over six years' observations for
M81-ULS1 despite the presence of apparent eclipse ingress/egress.
Indeed, the optical identification of the secondary with an AGB star implies an
orbital period of 30 years or longer if the AGB star overfills/underfills its
Roche lobe given its supergiant size ($600-1300 R_\odot$; Liu \& Di Stefano
2007).
Such a long orbital period is consistent with the absence of periodicity over
six years' observations.
On the other hand, it is possible to obtain folded light curves with only one
``eclipse'' for some periods scattered in the ranges of 30-35 days, if we
dismiss the two low-state ObsIDs 5940/5944.
However, there are no compelling reasons to dismiss these observations, and the
resulted periods are inconsistent with the long orbit period expected given the
AGB secondary.

% - we are puzzled by its "eclipses"

What are the origins for the dramatic flux drop/rise within an hour observed in
ObsIDs 390/5944?
The failure to detect any periodicity suggests that they are not eclipses from
the companion star blocking the primary; indeed, ingress/egress for an orbital
period of 30 years or longer is expected to last $\ge30$ days (or $1^\circ$),
three orders of magnitudes longer than observed in ObsIDs 390/5944.
If they originate from eclipsing by an intervening thick cloud of gas, this gas
must have a huge size orders of magnitudes larger than the X-ray emitting
region ($>10^4$ km), unbounded by the primary, move really fast at $\gg10$
km/sec, yet block the primary repetitively, and cause changes in $n_H$; we
cannot think of a reasonable scenario to produce such a cloud, and no
correlated changes in $n_H$ were observed.
This variability may be reminiscent of the stochastical oscillation between
high state and low state of some stellar mass black hole X-ray binaries (e.g.,
GRS 1915+105, Cyg X-1), but ULS1 has never shown the accompanied hard/soft
spectral state transition as observed for these sources.

Were ULS1 a white dwarf, such dramatic flux changes may be comparable to those
observed in other supersoft sources. 
For example, Nova V382 Vel showed a flux drop by a factor of 2 within 1.5 hours
(Orio et al. 2002), Nova Vl494 Aql exhibited a supersoft flare with flux
increase by a factor of 6 that lasted about 15 minutes (Drake et al. 2003), and
Nova V4743 Sagittarii showed the flux dropped to nearly zero within a few ks
(Ness et al. 2003).
Like in the case for M81-ULS1, the physical explanations are lacking for the
the dramatic variability in these sources.
Dramatic changes in the X-ray flux have also been observed, sometimes in
anti-correlation with the optical fluxes, for canonical supersoft sources such
as RX J0513.9-6951 (Reinsch et al. 1996) and CAL 83 (Greiner \& Di Stefano
2002).
They are usually understood as shifts of the spectrum peak due to the white
dwarf photosphere expansion and contraction as a result of accretion rate
changes, accompanied by the induced changes in the irradiation and heating of
the accretion disk.
The time scales for such changes, however, are a few days or longer rather than
hours as observed in M81-ULS1.

The short timescale variability, however, can in principle be explained by the
accretion disk changes if the accretor is an intermediate mass black hole.
In response to the variations in the accretion rate, the accretion disk
structure and temperature can change in as short as $10^3$ seconds even if the
response time is as long as $10^3\times$ the dynamical timescale, which is
short ($\le1$ second) for black hole accretors.
Because the accretion disk temperature is rather low ($\le100$ eV), the X-ray
observations are only catching the exponential tail of the emergent spectrum
from the accretion disk, and a small change in temperature can lead to dramatic
changes in the X-ray flux.
We note that this mechanism has been used to explain the dramatic X-ray
variability in narrow line Seyfert I nuclei and in some canonical supersoft
sources as mentioned above.

\acknowledgements

We would like to thank Rosanne Di Stefano, Jeff McClintock, Ramesh Narayan and
his Friday group meeting for helpful discussions.  JFL acknowledges the support for
this work provided by NASA through the Chandra Fellowship Program, grant
PF6-70043.

% figures

\begin{figure}

\plotone{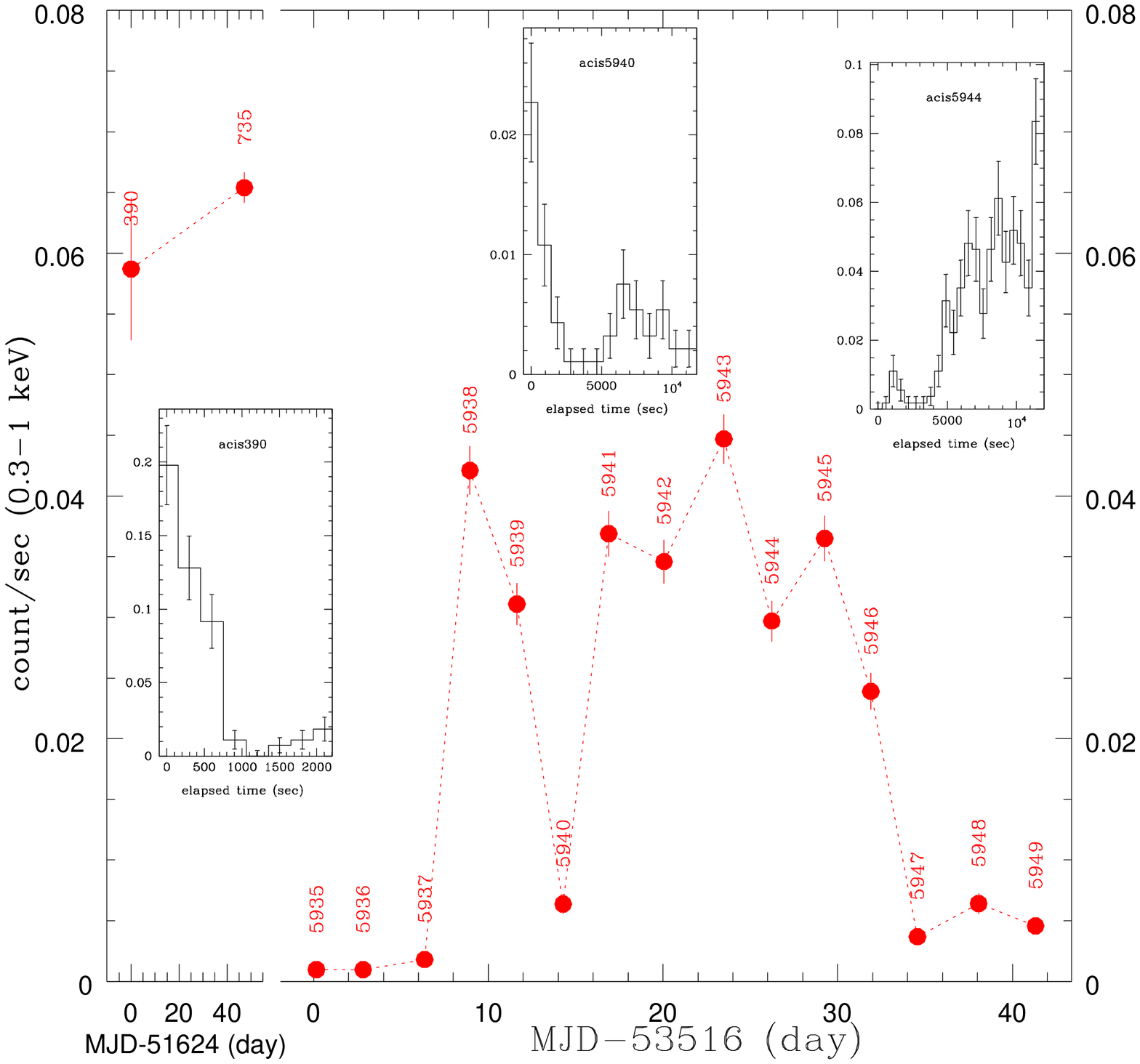}

\caption{Light curve for ULS1 from 17 Chandra observations over six years. ULS1
was on ACIS-S3 chip in all observations.  The ObsID is labeled for each
observation. The inserts show the binned light curves for three observations
with eclipse-like sudden drop and rise in less than one hour. The error bars
are calculated for Poisson statistics.  }

\end{figure}

\begin{figure}

\plotone{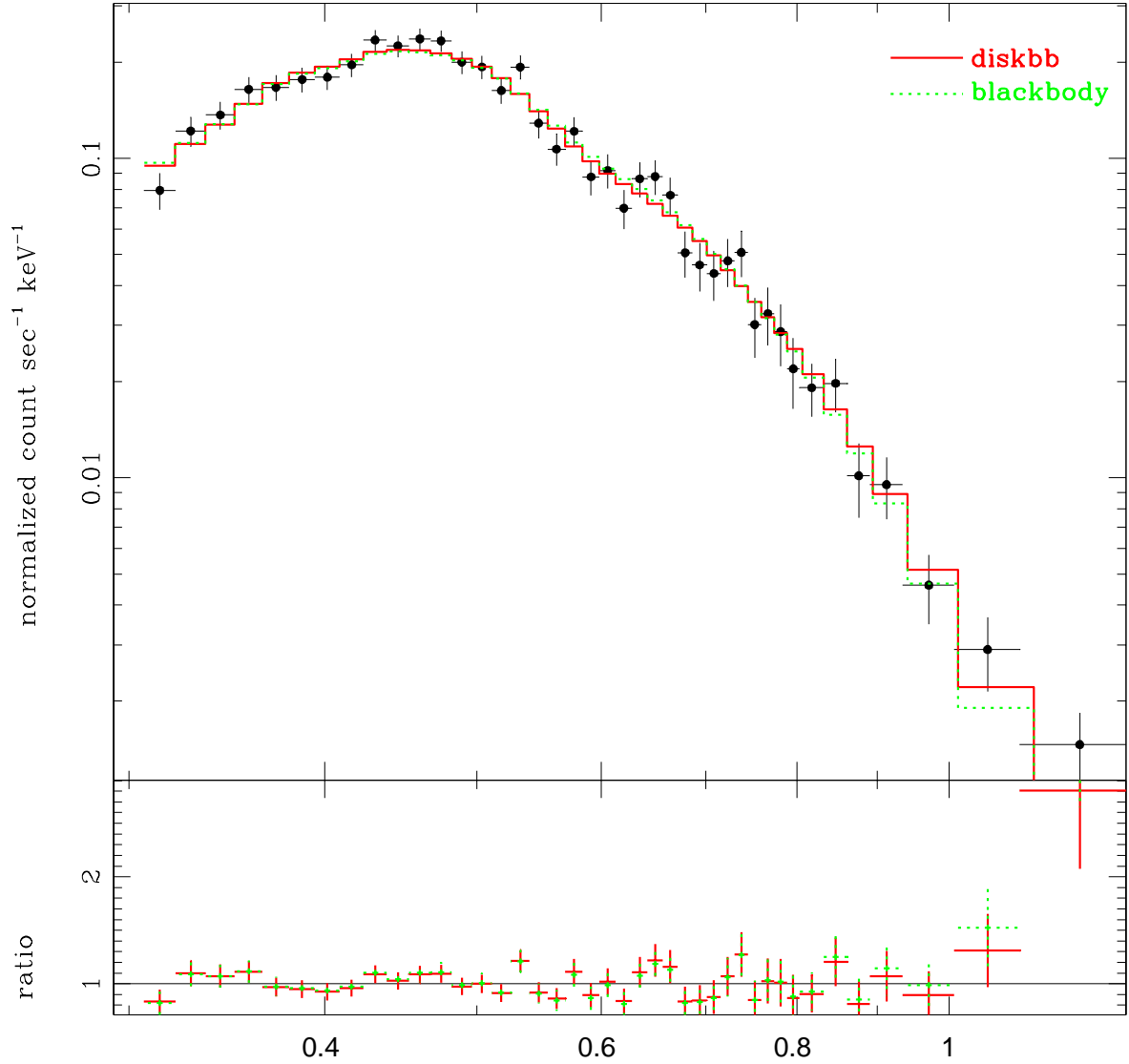}

\caption{ULS1 spectrum from ObsID 735 fitted by blackbody and diskbb models.
The difference between the blackbody model (green dotted line) and the diskbb
model (red solid line) only becomes discernible above 0.8 keV, with diskbb
predicting slightly more counts than blackbody. The fitting residuals above 1
keV suggest the presence of a hard component barely detected by ACIS.}

\end{figure}

\begin{figure}
\plotone{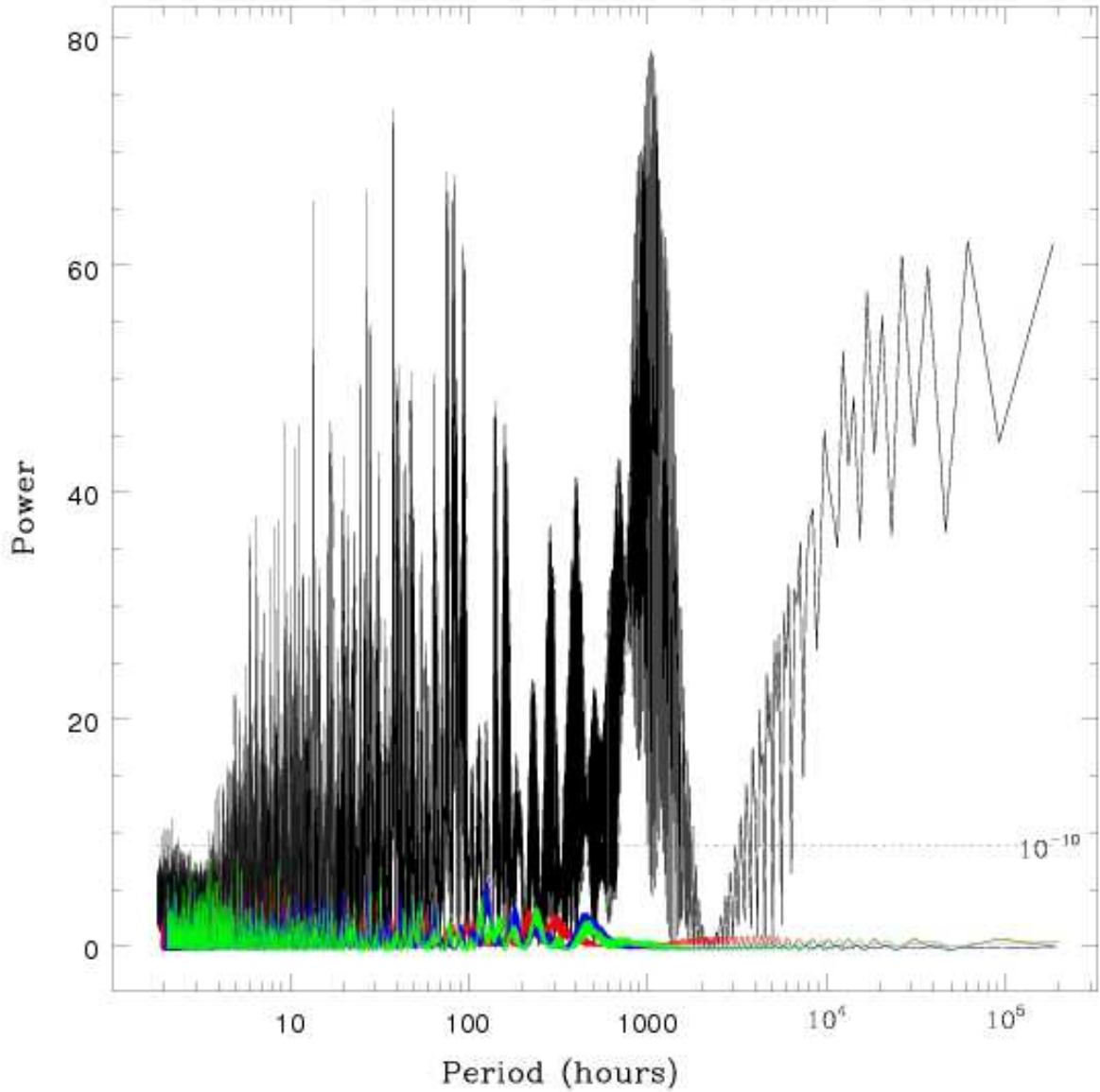}
\caption{Lomb periodogram for the 17 binned light curves of M81-ULS1.  The
dotted line indicates the power the probability for which to come from random
fluctuations is $10^{-10}$ in our calculations. Overlayed in color are the Lomb
periodograms for three Monte Carlo permutated data. }

\end{figure}

\begin{figure}
\plotone{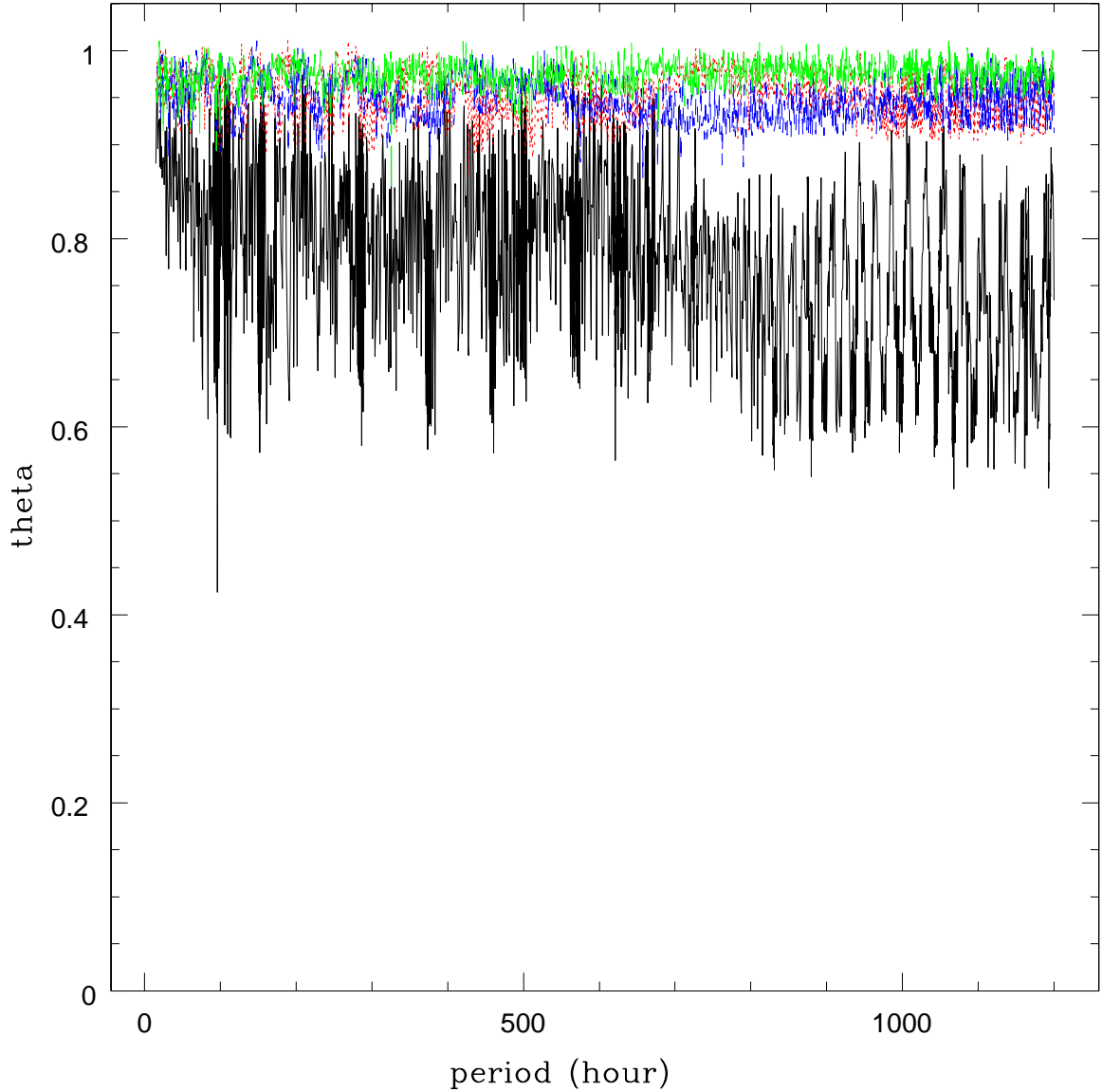}

\caption{The phase dispersion for trial periods from 15 hours to 50 days for
the 17 binned light curves of M81-ULS1. For our calculations, the probability
of $10^{-6}$ for a candidate period to come from random fluctuations
corresponds to the phase dispersion measure $\Theta$ of 0.66. The most
significant candidate period is 96.291 hours with $\Theta=0.424$ and the
probability of $10^{-21}$. Overlayed in colors are the phase dispersions for
three Monte Carlo permutated data.    }

\end{figure}

\begin{figure}

\plotone{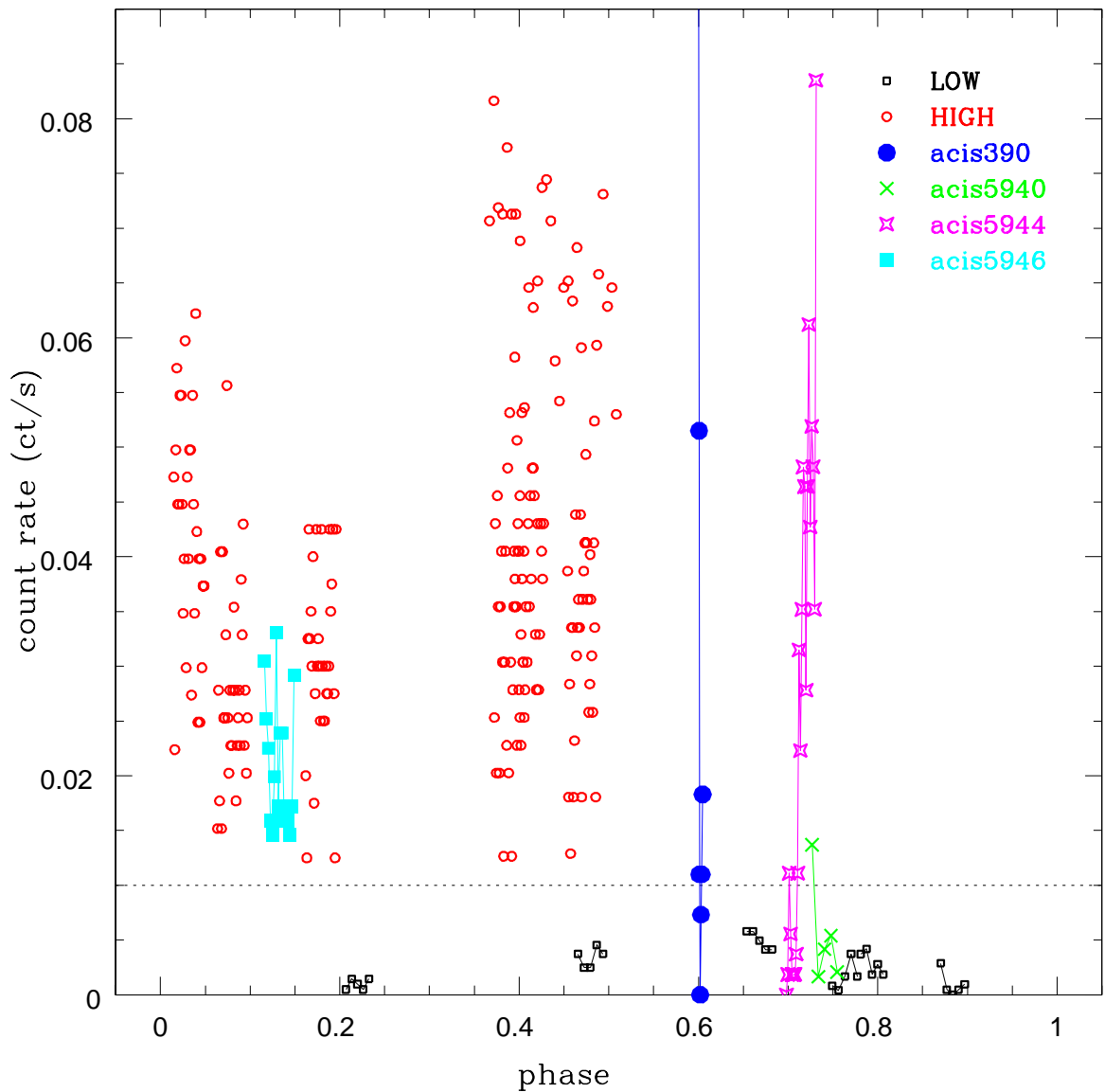}

\caption{Folded light curves for a candidate period of 96.291 hours. This
period was picked up with the phase dispersion minimization method. There are
four dips of low-state bins, or ``eclipses'', out of the high-state bins, at
phases 0.3, 0.55, 0.65 and 0.9, respectively. The dividing line between the
low-state bins and the high-state bins is the count rate of 0.01 as marked 
by the dotted line. This candidate period is not the true orbital period, since
a true period should give only one eclipse in the folded light curves. }

\end{figure}

\begin{figure}
\plotone{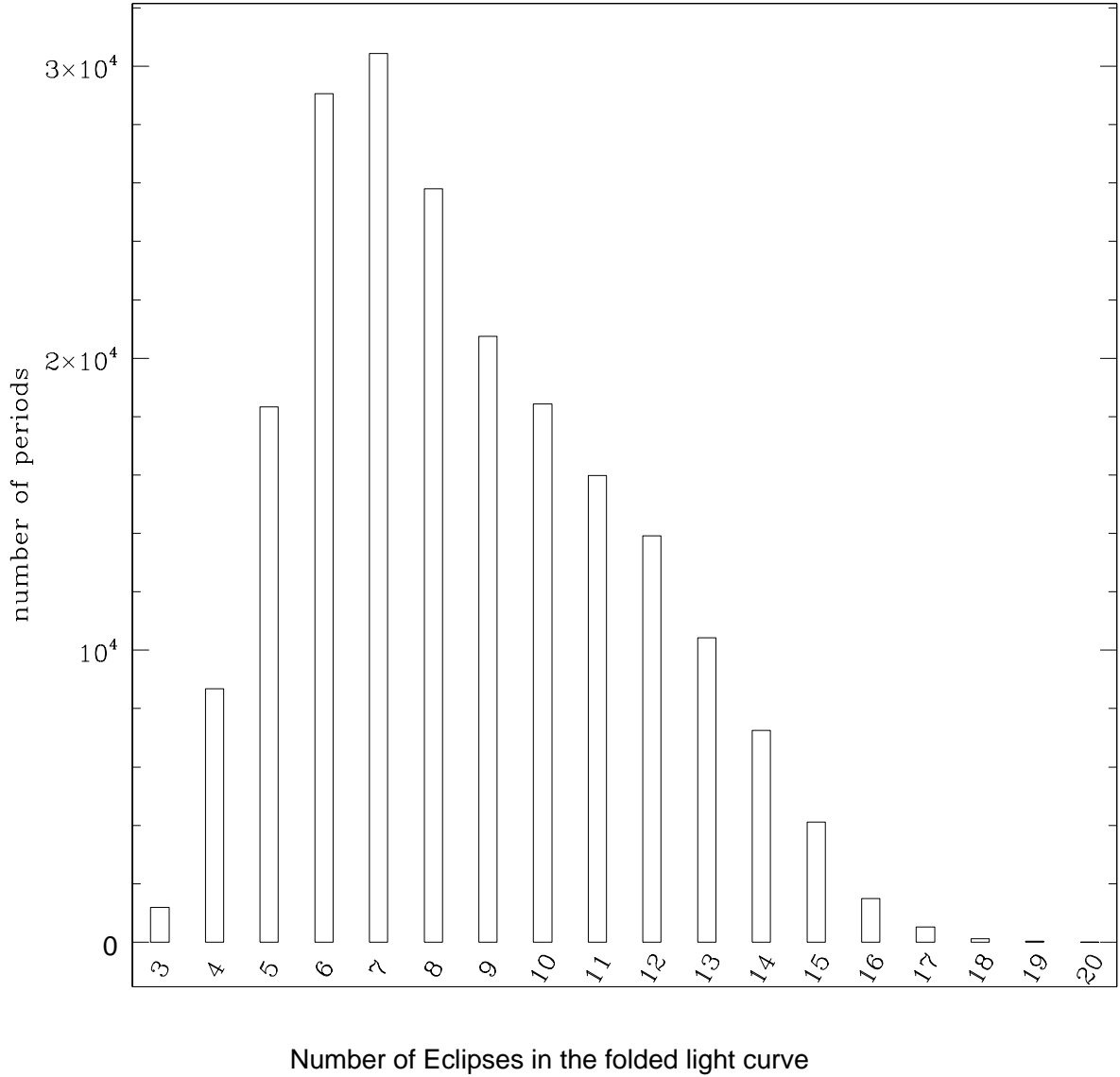}
\caption{Histogram for the number of eclipses in the folded light curves.
Under the discretization scheme described in the text, about $2\times10^5$
periods were tried for the continuous period range from 15 hours to 50 days
excluding the disallowed periods. All folded light curves have three or more
eclipses. }

\end{figure}

\begin{figure}

\plotone{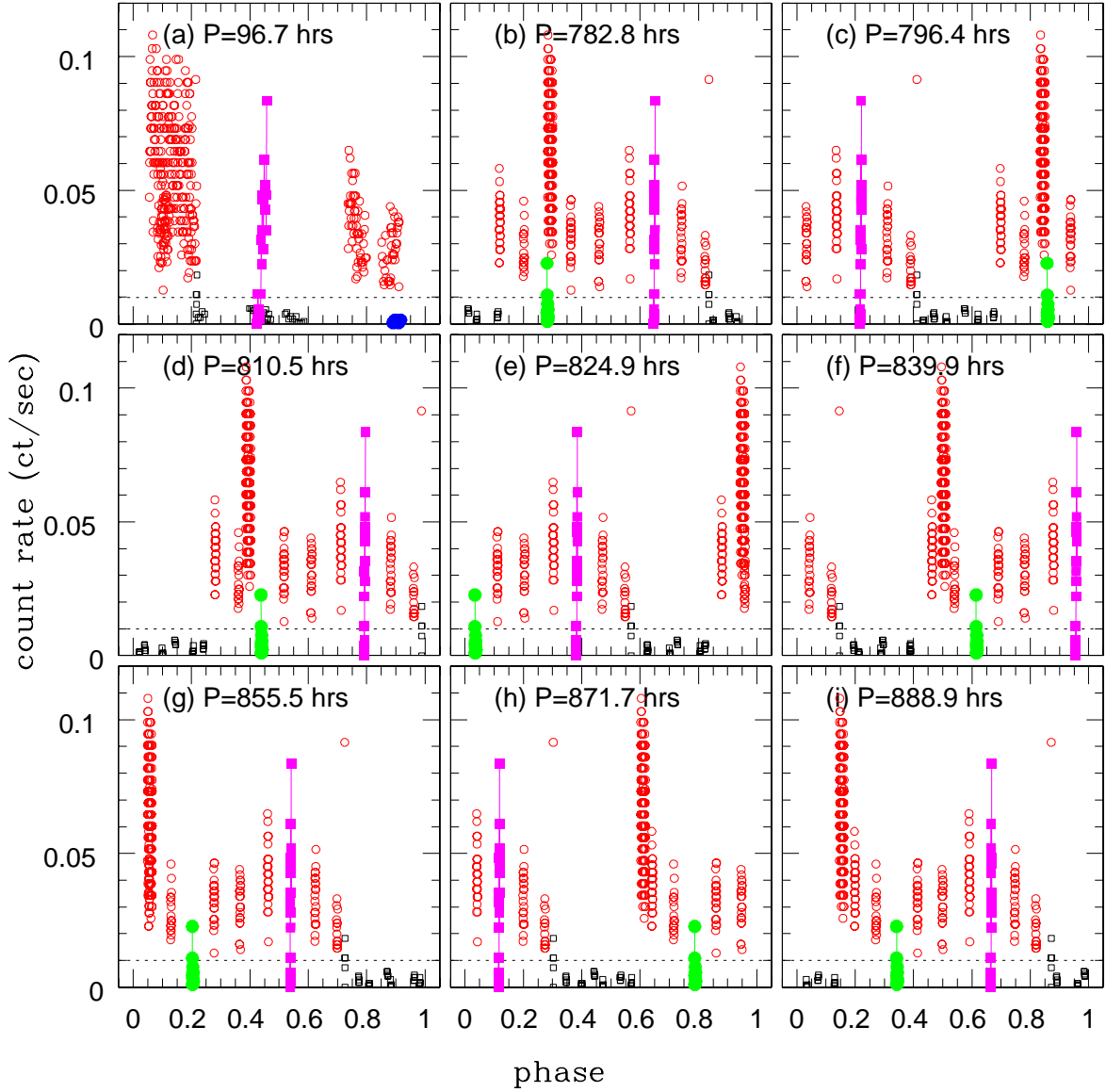}

\caption{Viable periods with only one ``eclipse'' in the folded light curves
after removal of two low-state observations.  Open circles are high-state bins
considered as out-of-eclipse, open squares are low-state bins considered as
in-eclipse. In panel (a), the two low-state observations ObsID 5935 (filled
circle) and ObsID 5944 (filled square) are considered as out-of-eclipse. In all
other panels, the two low-state observations ObsID 5940 (filled circle) and
ObsID 5944 are considered as out-of-eclipse. The dotted line in all panels
marks the count rate of 0.01 ct/s.}

\end{figure}

%\end{document}

% figures

% tables

\begin{deluxetable}{lccrrrrccccc}
\rotate
%\tabletypesize{\footnotesize}
\tabletypesize{\scriptsize}
\tablecaption{Chandra ACIS-S3 observations for M81-ULS1\tablenotemark{a}}
\tablehead{
 \colhead{ObsID} & \colhead{Obs-Date} & \colhead{OAA} & \colhead{VigF} & \colhead{ExpT}  & \colhead{Net} & \colhead{Soft} & \colhead{CountRate} & \colhead{Flux} & \colhead{$kT_{bb}$} & \colhead{$n_H$} & \colhead{$\chi_\nu^2/$dof} \\
 \colhead{} & \colhead{} & \colhead{($\prime$)} & \colhead{} & \colhead{(ksec)}  & \colhead{Counts} & \colhead{Counts} & \colhead{(ct/ksec)} & \colhead{ erg/s/cm$^2$ } & \colhead{(eV)} & \colhead{($10^{20}$ cm$^{-2}$)} & \colhead{} 
}

\startdata

 acis390   & 2000-03-21 & 2.5 & 0.91 &  2.4 &      127    &  107  & 58 $\pm$ 5.2  & 5.32e-13 &            &             &            \\
 acis735   & 2000-05-07 & 3.2 & 0.97 & 49.9 &     3182    & 3155  & 66 $\pm$ 1.2  & 5.28e-13 &  73 $\pm$ 1.5 & 8.6 $\pm$ 0.9 & 1.196/38      \\
 acis5935  & 2005-05-26 & 1.0 & 0.92 & 11.0 &       10    &    7  & 1.0 $\pm$ 0.3 & 9.23e-15 &            &             &            \\
 acis5936  & 2005-05-28 & 1.0 & 0.90 & 11.4 &       10    &    9  & 1.0 $\pm$ 0.3 & 9.39e-15 &            &             &            \\
 acis5937  & 2005-06-01 & 0.9 & 0.99 & 12.0 &       20    &   18  & 1.7 $\pm$ 0.4 & 1.53e-14 &            &             &            \\
 acis5938  & 2005-06-03 & 0.9 & 0.99 & 11.8 &      456    &  448  & 39 $\pm$ 1.8  & 3.77e-13 &  72 $\pm$ 5.8 & 20.6 $\pm$ 100 & 1.421/16   \\
 acis5939  & 2005-06-06 & 0.8 & 0.99 & 11.8 &      329    &  326  & 28 $\pm$ 1.6  & 2.82e-13 &  73 $\pm$ 7.4 & 9.5 $\pm$ 6.3 & 1.519/14      \\
 acis5940  & 2005-06-09 & 0.8 & 0.99 & 12.0 &       65    &   63  & 5.5 $\pm$ 0.7 & 5.55e-14 &            &             &            \\
 acis5941  & 2005-06-11 & 0.8 & 0.99 & 11.8 &      387    &  381  & 33 $\pm$ 1.7  & 2.98e-13 &  78 $\pm$ 5.9 & 17.6 $\pm$ 6.0 & 0.602/15     \\
 acis5942  & 2005-06-15 & 0.8 & 0.99 & 12.0 &      371    &  364  & 31 $\pm$ 1.6  & 3.19e-13 &  65 $\pm$ 4.4 & 17.7 $\pm$ 5.8 & 0.575/15     \\
 acis5943  & 2005-06-18 & 0.8 & 0.99 & 12.0 &      511    &  498  & 43 $\pm$ 1.9  & 3.77e-13 &  87 $\pm$ 5.5 & 11.2 $\pm$ 4.3 & 1.129/22     \\
 acis5944  & 2005-06-21 & 0.8 & 0.99 & 11.8 &      334    &  312  & 29 $\pm$ 1.6  & 2.78e-13 &  99 $\pm$ 8.9 & 19.3 $\pm$ 8.2 & 0.921/15    \\
 acis5945  & 2005-06-24 & 0.7 & 0.99 & 11.6 &      380    &  376  & 33 $\pm$ 1.7  & 2.90e-13 &  65 $\pm$ 4.8 & 17.8 $\pm$ 6.3 & 0.810/14    \\
 acis5946  & 2005-06-26 & 0.7 & 0.99 & 12.0 &      253    &  252  & 21 $\pm$ 1.3  & 2.06e-13 &  68 $\pm$ 5.8 & 9.2 $\pm$ 3.9 & 0.683/10      \\
 acis5947  & 2005-06-29 & 0.6 & 0.99 & 10.7 &       31    &   30  & 2.9 $\pm$ 0.5 & 2.50e-14 &            &             &            \\
 acis5948  & 2005-07-03 & 0.6 & 0.99 & 12.0 &       60    &   59  & 5.1 $\pm$ 0.7 & 4.80e-14 &            &             &            \\
 acis5949  & 2005-07-06 & 0.6 & 0.99 & 12.0 &       41    &   40  & 3.5 $\pm$ 0.5 & 2.99e-14 &            &             &            \\

\enddata 

\tablenotetext{a}{The columns are (1) ObsID, (2) MJD, (3) the chip M81-ULS1 was
on, (4) the off-axis angle in arcmin, (5) vignetting factor, (6) live time
after correction for deadtime, (7) net counts in 0.3-8.0 keV, (8) counts in the
soft band 0.3-1.1 keV, (9) count rates after vignetting correction, (10) the
0.3-8 keV flux computed from the corrected count rate assuming a power-law
spectrum with photon index 1.7, (11) the temperature in eV for the blackbody
fit to the spectrum in 0.3-2 keV, (12) the neutral hydrogen column density in
$10^{20}$ cm$^{-2}$, and (13) the reduced $\chi^2$ and degree of freedom for
the spectral fit. }

\end{deluxetable}

\end{document}